# Tuning Band Alignment and Optical Properties of 2D van der Waals Heterostructure via Ferroelectric Polarization Switching

Dimuthu Wijethunge[1,2], Lei Zhang[1,2], Cheng Tang[1,2] and Aijun Du[1,2]*

[1]*School of Chemistry and Physics, Queensland University of Technology, Gardens Point Campus, Brisbane, QLD 4000, Australia*

[2]*Centre for Materials Science, Queensland University of Technology, Gardens Point Campus, Brisbane, QLD 4000, Australia*

*\*aijun.du@qut.edu.au*

## Abstract

Favourable band alignment and excellent visible light response are vital for photochemical water splitting. In this work, we have theoretically investigated how ferroelectric polarization and its reversibility in direction can be utilized to modulate the band alignment and optical absorption properties. For this objective, 2D van der Waals heterostructures (HTSs) are constructed by interfacing monolayer $MoS_2$ with ferroelectric $In_2Se_3$. We find the switch of polarization direction has dramatically changed the band alignment, thus facilitating different type of reactions. In $In_2Se_3/MoS_2/In_2Se_3$ heterostructures, one polarization direction supports hydrogen evolution reaction and another polarization direction can favour oxygen evolution reaction. These can be used to create tuneable photocatalyst materials where water reduction reactions can be selectively controlled by polarization switching. The modulation of band alignment is attributed to the shift of reaction potential caused by spontaneous polarization. Additionally, the formed type-II van der Waals HTSs also significantly improve charge separation and enhance the optical absorption in the visible and infrared regions. Our results pave a way in the design of van der Waals HTSs for water splitting using ferroelectric materials.

**Keywords:** Photocatalyst, Ferroelectric, $MoS_2$, $In_2Se_3$, Heterostructures, Water splitting, 2D materials

# 1. Introduction

In the context of searching sustainable, economical, and efficient energy generation processes, nature provides perfect insight on what future energy generation should be like. In photosynthesis process, solar energy is converted into chemical energy by plants and such type of conversion can be achieved also through photocatalyst materials, even though the efficiency is far below the natural photosynthesis process. Photocatalyst material has gained immense popularity in research community since the first discovery of ZnO in 1911[1]. Later, another popular photocatalyst, $TiO_2$, was reported in 1938, through production of active oxides by absorbing UV on the surface[2]. After several decades fujishima et al have introduced electrochemical photolysis of water using $TiO_2$ and Pt electrodes[3]. Since then water splitting through solar irradiation into hydrogen and oxygen identified as the efficient reaction for solar chemical energy conversion. Later various photocatalyst such as $CeO_2$[4], $BiVO_4$ [5], $WO_3$[6], $CdS$[7], $GaN$[8], $WSe_2$[9] and $InGaP$[10] were reported. Among them $BiVO_4$ based photoanode has demonstrated the highest conversion efficiencies around 5.2 % [11]. With the discovery of graphene, the use of two-dimensional (2D) materials as a photocatalyst become major focus owing to their unique characteristics such as ideal band gap, and high surface area. 2D materials such as $MoS_2$[12-13], $SiP$[14], $g$-$C_3N_4$[15], $InSe$[16], $PdSeO_3$[17] and $In_2Se_3$[18] have shown good photocatalytic properties for water splitting.

Theoretically, a perfect efficient photocatalyst should possess a bandgap higher than 1.23 eV but not greater than 3.0 eV. Here 1.23 eV is the free energy of water splitting and band gap should be less than 3.0 eV for efficient absorbance of the solar energy [19-20]. In addition, band edges of photocatalyst material should be aligned with potentials of hydrogen evolution reaction (HER) and oxygen evolution reaction (OER). Conduction Band Minimum (CBM) should be higher than -4.44 eV (pH=0 for efficient HER reaction [21]. Valence Band Minimum (VBM) should be lower than -5.67eV (pH=0), which is the potential for OER reaction[21]. Equation (1) and (2) illustrate the OER and HER where electrons and holes are denoted by e and h, respectively. Additionally, e and h must be well separated. In general, it is extremely challenging to find an efficient photocatalytic material which satisfy all these requirements

$$H_2O + 2h^+ \longrightarrow 2H^+ + 1/2O_2 \quad (1)$$

$$2H^+ + 2e^- \longrightarrow H_2 \quad (2)$$

Ferroelectric material exhibits spontaneous polarization in the absence of external electric field. Some research works have demonstrated influence of changing polarization direction on catalytic surface absorption and desorption[22], Schottky tunnel barrier[23], e-h charge separation[24] and bandstructure engineering[25]. Recently few 2D ferroelectric materials with out-of-plane polarization have been discovered and synthesized[26-28]. One interesting question is to investigate possibility of improving the photocatalysis performance by forming Van der Waals Heterostructures (HTSs) with ferroelectric materials. It is highly expected that ferroelectric materials with spontaneous polarization and reversibility of the polarization direction can significantly modulate the band alignment. Additionally, by forming Type-II Van der Waals HTSs with ferroelectric materials, it can also facilitate charge separation and enhance optical absorption in photocatalyst[29]. 2D α-$In_2Se_3$ with a band gap of 1.3 eV is an experimentally realized ferroelectric material that have demonstrated potential for water splitting applications[18, 30]. 2D $In_2Se_3$ possesses considerable out-of-plane ferroelectric polarization and the band gap can be reduced when it forms a vertical HTS with $WSe_2$ monolayers[31]. $In_2Se_3$/$MoS_2$ HTS has been experimentally synthesized and used as a photoanode in water splitting[32]. Optical absorption properties of $In_2Se_3$/$MoS_2$ HTS has been improved compared to intrinsic $In_2Se_3$ and $MoS_2$[33].

2D $MoS_2$ has a bandgap around 1.8 eV[34] which is larger than 1.23eV and its band alignment support both HER and OER [12]. $MoS_2$ has good carrier mobility and high surface to volume ratio but charge recombination is much higher considering the amount of charge generated. It also has the leverage to reduce from its original band gap of 1.8 eV to 1.23 eV, which automatically increases the light absorption at infrared wavelength. Studies have shown, constructing HTS with certain materials can reduce both band gap and charge recombination in $MoS_2$ [35-37]. In HTSs, charge recombination is reduced greatly when CBM and VBM appears in two different layers.

In this work, we have constructed HTSs using 2D $MoS_2$ and ferroelectric $In_2Se_3$ and found that band position can only support HER or OER depending on the ferroelectric polarization direction. When $MoS_2$ layer is sandwiched between top and bottom $In_2Se_3$ to form $In_2Se_3$/$MoS_2$/$In_2Se_3$ HTS, three distinct polarization states can be produced and each exhibit different band alignment with water reduction potential. One polarization state facilitates both HER and OER and the other polarizations states only facilitate either OER or HER. This allows us to shift the water reduction potentials in photocatalyst material simply by switching ferroelectric polarization in ferroelectric materials, potentially enhance the photocatalysis efficiency in water splitting. Additionally, the formed type-II van der Waals HTSs also significantly improve charge separation and enhance the optical absorption in the visible and infrared regions.

## 2. Methodology

All calculations were conducted by using density functional theory (DFT) with the assistance of Vienna Ab Initio Simulation Package (VASP)[38]. Core and valence electrons were interpreted using projected-wave (PAW) argument method[39-40]. Both Perdew, Burke, and Ernzerh (PBE) method[41] and Heyd-Scuseria-Ernzerhof (HSE) method[42] were used in the calculation of bandstructure. DFT-D3 methods was used to incorporate long range van der Waals interactions in HTSs[43]. Vacuum region of 20Å was maintained to prevent interaction between periodic images. Cut off energy of 500eV was used and the convergence criteria of forces on the atoms were set to -0.001eV/ Å and energy convergence criteria was set to 1E-$^{06}$. 5×5×1 and 7×7×1 gamma centred k-point meshes were used for the geometry relaxations and electron structure calculations, respectively.

2D $In_2Se_3$ and $MoS_2$ crystal structures were illustrated in Figure 1(a)-(d). Table 1 shows the calculated lattice parameters for $In_2Se_3$ and $MoS_2$ and those are in good agreement with the experimental values. First, 2D $In_2Se_3$/$MoS_2$ HTSs were constructed using 3×3 $In_2Se_3$ and 4×4 $MoS_2$ unit cells to achieve minimum lattice mismatch (3%). Two different polarization states (type-A and type-B) for 2D $In_2Se_3$/$MoS_2$ HTSs were illustrated in Figure 1(e) and (f). Next, $In_2Se_3$/$MoS_2$/$In_2Se_3$ HTSs were constructed as shown in Figure 1(g)-(i) by sandwiching $MoS_2$ between $In_2Se_3$ layers. Based on polarization direction, three different polarizations states can be found and they were denoted as type-BB, type-AA and type-AB. Reversing of the polarization occurs through rearrangement of the contacted surfaces (A and B). Naming of the HTSs are based on which side of $In_2Se_3$ is in contact with $MoS_2$. For example, HTS is named as type-AA, when the side A of $In_2Se_3$ layers are in contact with $MoS_2$ from both top and bottom sides.

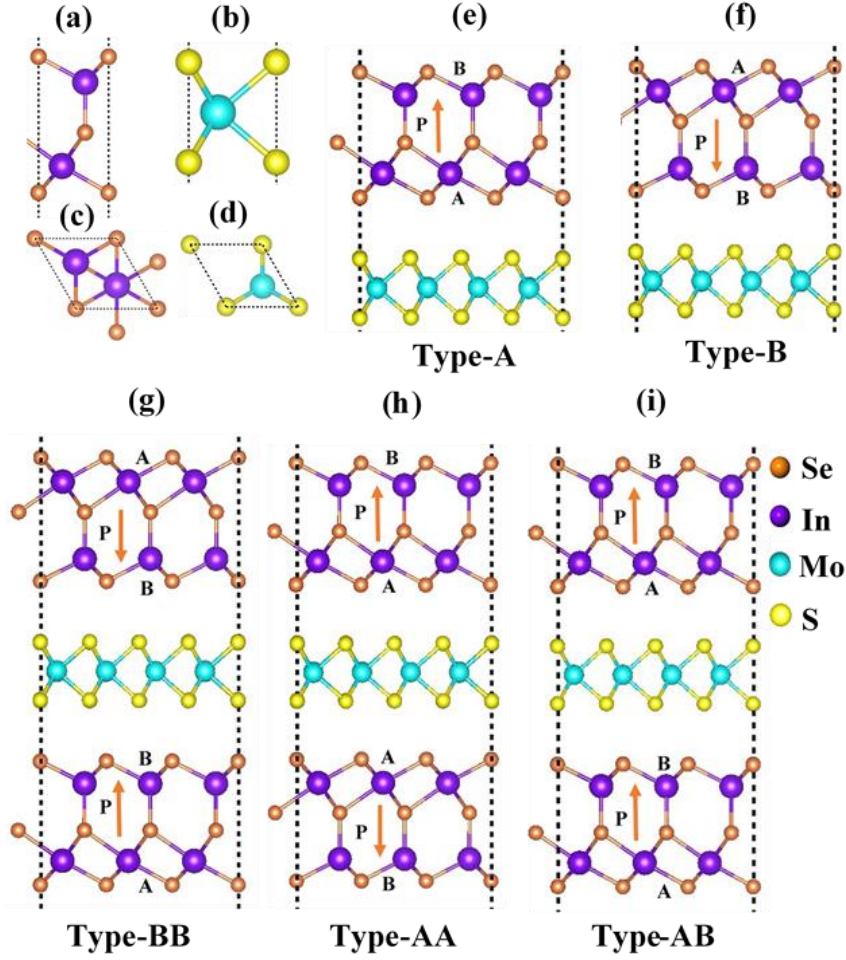

***Figure 1*** *:Side view of (a) $In_2Se_3$ (b) $MoS_2$ and top view of (c) $In_2Se_3$ (d) $MoS_2$ monolayers. Side view of (e) type-A (f) type-B, $In_2Se_3/MoS_2$ Heterostructures and (g) type-BB (h) type-AA (i) type-AB, $In_2Se_3/MoS_2/In_2Se_3$, Heterostructures. Polarization direction is indicated by the arrow.*

***Table 1:*** *Lattice parameters of Monolayer $In_2Se_3$ and $MoS_2$ unit cells, compared with reference values.*

|  | a,b(Å) this work | c(Å) this work | Ref a,b(Å) | Ref c(Å) |
|---|---|---|---|---|
| $In_2Se_3$ | 4.07 | 6.84 | $4.10^{31}, 4.11^{18}$ | $6.8^{31}$ |
| $MoS_2$ | 3.16 | 3.13 | $3.22^{44}, 3.18^{12}$ | $3.17^{45}$ |

## 3. Results and Discussion
### 3.1. Binding Energy

Bandstructure of the intrinsic $In_2Se_3$ and $MoS_2$ was calculated using both HSE and PBE methods. As given in the Table 2, our calculated values have good agreement to both experimental and previous theoretical calculations. According to the results, $MoS_2$ bandgap was accurate when calculated by PBE method, while bandgap of $In_2Se_3$ was accurate when calculated using HSE method.

*Table 2: Bandgap of monolayer $In_2Se_3$ and $MoS_2$ which calculated using PBE and HSE methods and compared with values obtained from reference calculations done by PBE, HSE and experimental methods.*

|  | PBE, this work | HSE, this work | PBE | HSE | Experiment |
| --- | --- | --- | --- | --- | --- |
| QL $In_2Se_3$ | 0.82 eV | 1.51 eV | 0.78eV[31], 0.77eV[46] | 1.46 eV[31] | 1.3 eV[30] |
| Monolayer $MoS_2$ | 1.75 eV | 2.07 eV | 1.9 eV[47], 1.78 eV[48], 1.70 eV[49] | 2.01 eV[50] | 1.8 eV[34] |

Binding energies ($E_b$) of all HTSs were calculated using the Equation 3 to check the stability of HTSs. Here $E_{tot}$, $E_{In2Se3}$ and $E_{MoS2}$ are the total energy of the system, energy of $In_2Se_3$ layer and energy of $MoS_2$ layer, respectively. For the HTSs with single layer of $In_2Se_3$, one $E_{In2Se3}$ term can be neglected. Binding energies of all HTS having negative values as indicated in Table 3, suggesting, HTSs are stable and energy wise favourable to form.

$E_b = E_{tot} - E_{In2Se3(top)} - E_{MoS2} - E_{In2Se3(bottom)}$ (3)

*Table 3: Calculated binding energies of all heterostructures*

|  | $E_{tot}$ (eV) | $E_{In2Se3}$ (eV) | $E_{MoS2}$ (eV) | $E_b$ (eV) |
| --- | --- | --- | --- | --- |
| $In_2Se_3$/$MoS_2$ type-A | -529.584 | -170.907 | -356.628 | -2.049 |
| $In_2Se_3$/$MoS_2$ type-B | -529.428 | -170.907 | -356.628 | -1.893 |
| $In_2Se_3$/$MoS_2$/$In_2Se_3$ type-AB | -702.985 | -170.907 | -356.628 | -4.543 |
| $In_2Se_3$/$MoS_2$/$In_2Se_3$ type-AA | -702.955 | -170.907 | -356.628 | -4.513 |
| $In_2Se_3$/$MoS_2$/$In_2Se_3$ type-BB | -702.732 | -170.907 | -356.628 | -4.290 |

### 3.2. $In_2Se_3$/$MoS_2$ Heterostructures

Bandstructure and contribution of each layer to its bands have been investigated in $In_2Se_3$/$MoS_2$ HTS as illustrated in the Figure 2. Type-A HTS shows band gap of 0.66 eV (PBE method) which is a reduced bandgap compared to intrinsic $In_2Se_3$ and $MoS_2$. On the other hand, type-B HTS has band gap of 0.84 eV (PBE method) which is similar to the band gap of $In_2Se_3$. According to the bandstructure of type-A HTS, valence band is contributed by $MoS_2$ layer and conduction band is contributed by $In_2Se_3$ layer. Hence charge recombination is greatly reduced owing to transferring of holes and electrons into two different layers. But in type-B configuration, both conduction and valence bands are mainly contributed by the $In_2Se_3$ layer, thus charge separation is not effectively facilitated. According to research, intrinsic $In_2Se_3$ also has better charge separation compared to $MoS_2$ monolayer due to its inbuilt polarization[18].

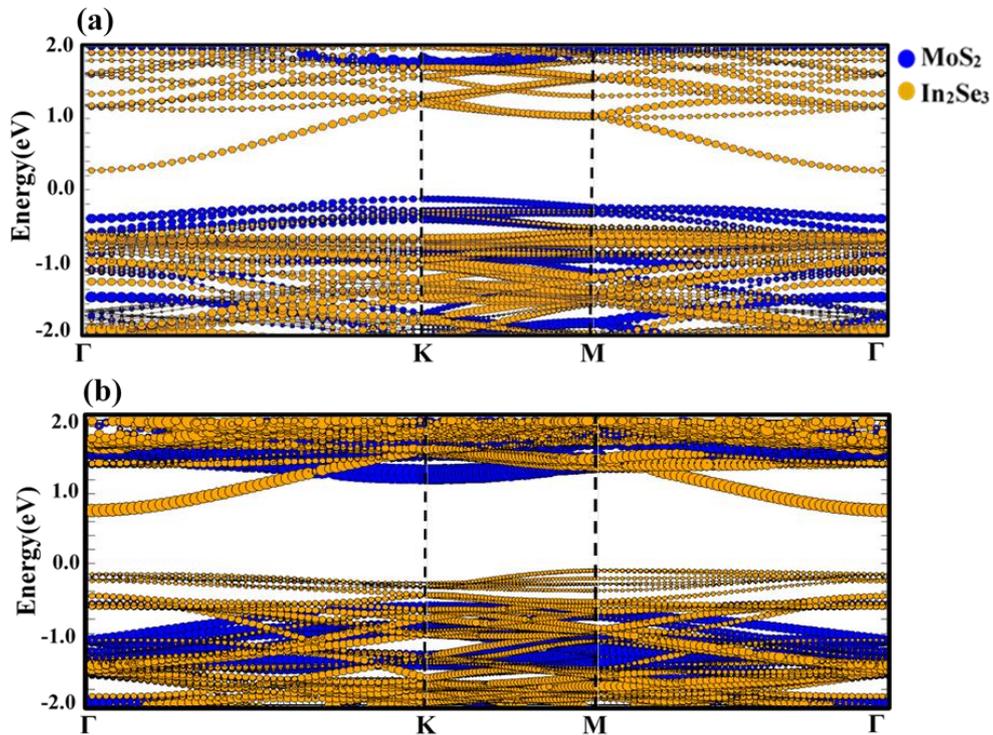

*Figure 2* : *Electronic Bandstructure of (a) type A (b) type B $In_2Se_3$/$MoS_2$ Heterostructure, calculated using PBE method. Contribution to each band from $MoS_2$ and $In_2Se_3$ layers are indicated in different colours.*

Figure 3 (a) and (b) illustrate the electrostatic potential (eV) of both HTSs and it show different vacuum levels at each end owing to the inbuilt polarization. Figure 3(c) and (d) show the band alignment of type-A and type-B configurations. Type-A have CBM above the HER potential and VBM below the OER potential, thus, it supports both reactions. However, in type-B configuration, VBM is higher than the OER potential and CBM is below the HER potential. Therefore, only HER is supported in type-B configuration. Results exhibit change of polarization direction has strong impact on band alignment. Through close observation, it can be seen that spontaneous polarization also plays critical role in the context of band aligning. In type-A HTS, CBM is -5.42 eV and yet it's above the HER potential. This happens as a result of different vacuum levels at both ends. Only $In_2Se_3$ have contributed to the CBM and vacuum level is lower at the end of $In_2Se_3$. Therefore, when considering HER potential respect to the vacuum level at the side of $In_2Se_3$, it becomes lower than the CBM. On the other hand, in type-B HTS, both CBM and VBM occurs in $In_2Se_3$ layer, which has the greater vacuum level. Therefore, the both potential values taken respect to same vacuum level. This feature is unique to polarized materials and it can assist better band alignment in photocatalyst materials.

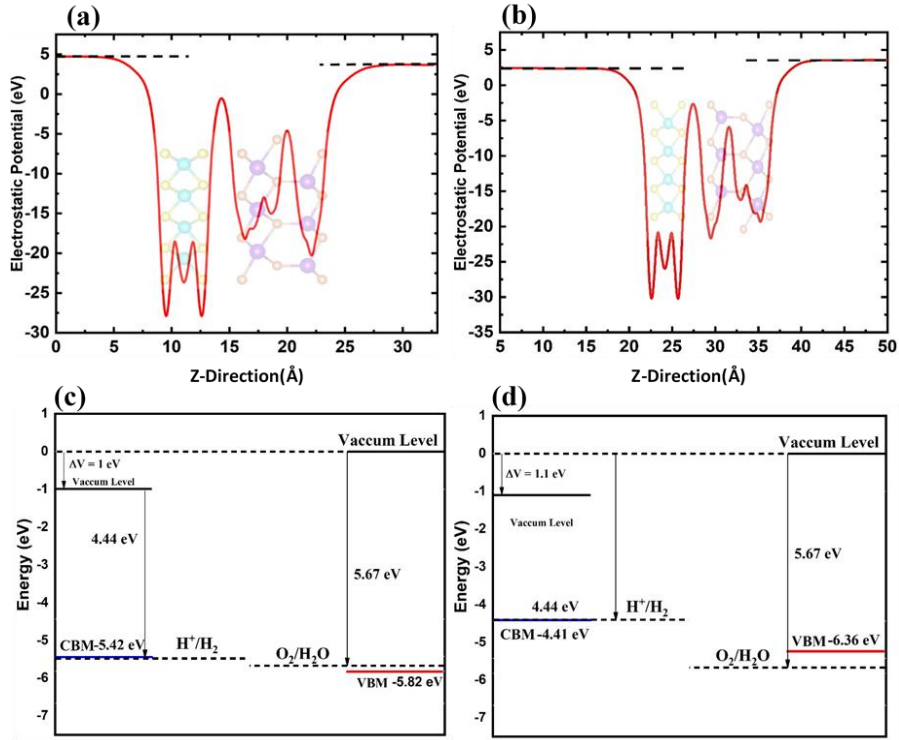

***Figure 3***: *Electrostatic potential of (a) type-A (b) type-B HTSs and alignment of VBM and CBM respect to HER and OER potentials of (c) type-A (d) type-B HTSs, calculated based on PBE method.*

### 3.3. $In_2Se_3/MoS_2/In_2Se_3$ Heterostructures

Here, HTSs were constructed by sandwiching the $MoS_2$ layer between $In_2Se_3$ layers. In this arrangement, polarization becomes stronger and its controllability can be improved. Based on polarization direction, three states, namely, AA, BB and AB can be identified. In type-AA and type-BB HTSs, $In_2Se_3$ layers have polarization directions opposing to each other, hence HTSs have zero net spontaneous polarization. In AB HTS, polarization direction of the two layers are in same direction, thus, it's subjected to very high spontaneous polarization. Figure 4 shows the band structure and contribution to bands from each layer. In type-AA, CBM is contributed by both top and bottom $In_2Se_3$ layers and VBM is by the $MoS_2$. In type-BB, both CBM and VBM occur in both $In_2Se_3$ layers. In type-AB CBM occurs in top $In_2Se_3$ layer and VBM occurs in bottom $In_2Se_3$ layer. Therefore, type-AA and type-AB types have good charge separation while charge separation hasn't improved in type-BB.

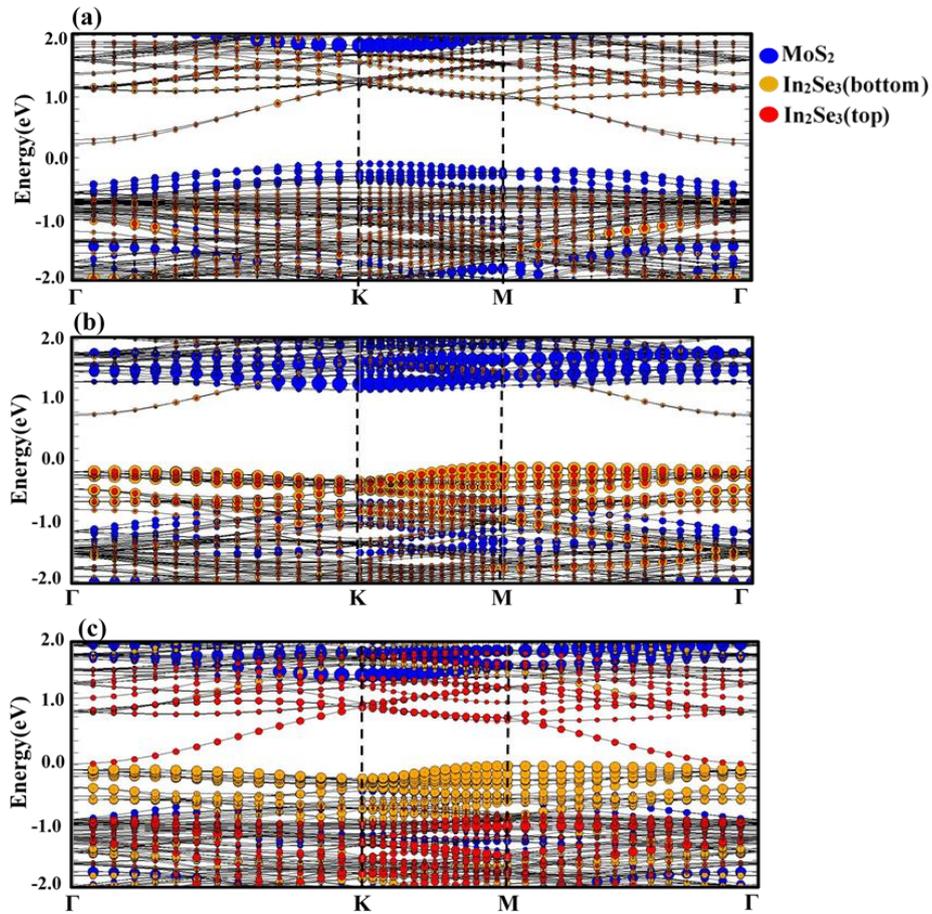

*Figure 4*: Electronic Bandstructure of (a) type-AA (b) type-BB (c) type-AB $In_2Se_3/MoS_2/In_2Se_3$ Heterostructure, calculated using PBE method. Contributions to each band from $MoS_2$, $In_2Se_3$ bottom layer and $In_2Se_3$ top layer are indicated in different colours.

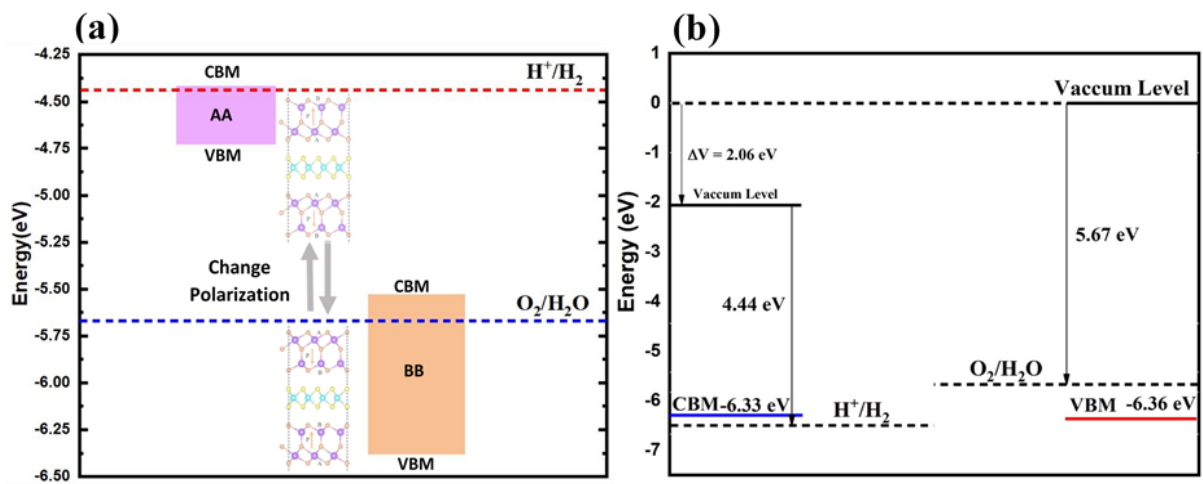

*Figure 5*: Alignment of VBM and CBM respect to HER and OER potentials of (a) type AA, type BB and (b) type AB HTSs which calculated using PBE method.

As illustrated in the Figure 5, type-AA configuration supports only HER and type-BB configuration supports only OER. But Type-AB configuration supports both reactions owing to assistance from existing inbuilt spontaneous polarization. This dramatic change of band alignment respect to the direction of polarization can lead to tuneable photocatalyst materials which its selectivity and efficiency of the chemical reactions can be controlled by changing the polarization direction.

Finally, optical absorption coefficient of all HTSs, $MoS_2$ and $In_2Se_3$ have been studied as shown in Figure 6. All HTSs except type-B HTS have shown increased optical absorption coefficient compared to $In_2Se_3$ and $MoS_2$ around the ultraviolet (UV) region. In visual light region, all HTSs, $MoS_2$ and $In_2Se_3$ have shown similar absorption coefficient values. As highlighted in Figure 6(b), optical absorption at infrared region also have shown considerable increase in all HTSs compared to $MoS_2$. The increase of infrared light absorption occurs due to reduction of bandgap in HTSs.

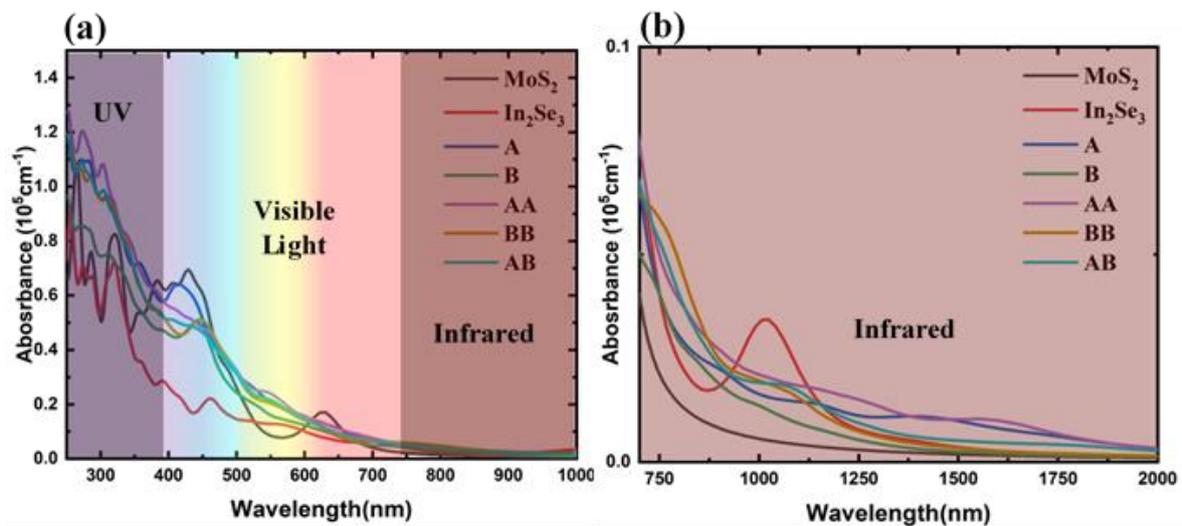

*Figure 6*: Optical absorption spectra in ( $MoS_2$ monolayer, $In_2Se_3$ monolayer, type-A, type-B, type-AA, type-BB and type-AB heterostructure represented by black, red, blue, green, violet, orange and cyan lines, respectively) (a) UV, Visible light and Infrared wavelengths (b) Extended infrared wavelengths, calculated based on PBE calculations.

4. Conclusion

In this work $In_2Se_3/MoS_2$ heterostructures with two polarization states, namely, type-A and type-B and $In_2Se_3/MoS_2/In_2Se_3$ heterostructures with three polarization states, namely, type-AA, type-BB and type-AB have been studied. Type-A and Type-AB HTSs support both OER and HER. Type-B and type-AA HTSs only support HER while Type-BB HTS only supports OER. Through HTSs formation, charge separation can be enhanced in photocatalyst materials as reflected by type-A, type-AA and type-AB configurations. All HTSs have shown increased optical absorption coefficient at UV and Infrared region of the solar spectrum compared to monolayer $MoS_2$ and $In_2Se_3$.

The ability to change band alignment through altering the polarization direction is one of the important findings of this work. This feature is evident in type-AA, type-BB and type-AB HTSs which their band alignments were drastically changed respective to the polarization direction. Applying this principle, external electric field can be used to reverse the polarization and change the material properties from photoanode to photocathode or vice versa. This also allows to change reaction paths according to the concentration of the products generated in the solution. If the products of one reaction is greater than the other, then the polarization direction can be changed to facilitate the other reaction using the same photocatalyst material. Interestingly, fast switching between $In_2Se_3/MoS_2/In_2Se_3$ three polarization states (AA, AB, BB) perhaps can provide an interesting result in water splitting applications. Another important finding is how the spontaneous polarization have shifted OER or HER potential values as shown in type-A and type-AB HTS configurations to facilitate the band alignment. Overall results of this work provide valuable insight on how ferroelectric materials can be used in photocatalyst application to improve their performance and may lead to development of next generation tuneable photocatalyst for water splitting.

## Acknowledgements

We highly acknowledge Queensland University of Technology (QUT) and National Computational Infrastructure (NCI) Australia for providing high performance computing (HPC) facilities to undertake this project.